# Electrical and optical properties of hydrated amorphous vanadium oxide

A. Pergament, A.Velichko[1], V. Putrolaynen, G. Stefanovich, N. Kuldin, A. Cheremisin, I. Feklistov, N. Khomlyuk

*Department of Physics, Petrozavodsk State University, 185910, Petrozavodsk, Russia*

[1]Authors to whom any correspondence should be addressed.

E-mail: velichko@psu.karelia.ru

**Abstract**

Electrical and optical properties of amorphous vanadium oxide thin films obtained by electrochemical anodic oxidation are studied. It is shown that under cathodic polarization the hydrogen insertion into vanadium oxide from an electrolyte occurs. Metal-insulator transition in amorphous $H_xVO_2$ is found to be preserved up to high concentration ($x \sim 1.5$) of hydrogen. Memory switching with the N-type negative differential resistance, associated with the $H^+$ ionic transfer, is observed in "V/hydrated amorphous vanadium oxide/Au" sandwich structures.

**Keywords:** vanadium oxide, hydrogen insertion, metal-insulator transition, switching, memory

**PACS numbers:** 71.30.+h, 72.80.Ga, 73.50.Fq, 61.43.-j

## 1. Introduction

A large variety of properties of hydrated oxides and hydrogen oxide bronzes (especially those of transition metal oxides [1]), in combination with their promising potential technical applications, has stimulated interest in electrochemical insertion of hydrogen into oxide networks. As an example, one can mention the studies of electrochromic and photochromic effects in transition metal oxides, such as $WO_3$, $V_2O_5$, and $MoO_3$ [2-4], which are associated with a transformation of electronic spectra due to either $H^+$ insertion/extraction from an electrolyte [2] or its redistribution inside the hydrated oxide [5]. Another example of the hydrogen-induced modification of the transition-metal oxide electronic structure is connected with vanadium dioxide, exhibiting the metal-insulator transition (MIT) at $T_t = 340$ K [1]. Hydration of crystalline $VO_2$ has been found to influence the parameters of the MIT in this compound leading, particularly, to a shift of the transition temperature $T_t$ toward the low-temperature region [6, 7]. In this work we have studied the effect of hydrogen insertion on the properties of thin-film amorphous vanadium oxides obtained by anodic oxidation.

## 2. Experimental

Vanadium oxide films were prepared by means of anodic oxidation [8] of vanadium metal layers deposited onto Si and glass-ceramic substrates by magnetron sputtering. X-ray diffraction studies showed that the films represented structurally disordered objects, and the phase compositions corresponded to either $VO_2$ or a mixture of $VO_2$ and $V_2O_5$, depending on the anodizing conditions [8, 9]; the film thickness $d$ was controlled by anodic potential. The process of oxidation was performed in the electrolyte containing 22 g of benzoic acid ($C_6H_5COOH$) and 40 ml of a saturated aqueous solution of sodium tetraborate ($Na_2B_4O_7 \times 10H_2O$) per one litre of acetone. The hydrogen insertion into the prepared films was carried out in that same electrolyte, but under cathodic polarization.

Current-voltage *I-V* characteristics of "V/hydrated amorphous vanadium oxide/Au" sandwich structures were recorded by a two-probe method with a Keithley 2410 Sourcemeter and a probe station SUSS PM5. The voltage scan rate was 2 V/sec. Top Au electrodes were deposited by thermal evaporation onto oxide film surfaces.

Optical measurements were carried out using a spectrophotometer SF-46 in the wavelength range $\lambda$ from 200 to 1100 nm.

## 3. Results and discussion

The optical reflectance spectrum of the as-prepared $VO_2$ anodic oxide film is presented in figure 1(a), curve 1. For the sample subject to cathodic polarization, a shift of the spectrum toward shorter wavelengths, accompanied by a corresponding colour change, is observed (curve 2). The occurrence of the MIT in the anodic $VO_2$ films is indicated by an abrupt and reversible change in the optical properties at $T \sim 330$ K (figure 1(b)).

This temperature practically coincides with the transition temperature for crystalline $VO_2$, $T_t = 340$ K, which evidences the fact that the MIT in $VO_2$ is preserved in the absence of the long-range crystallographic order and that electron-electron correlations play an important role in the transition mechanism [8-10]. The structural disorder slightly lowers the $T_t$ (from 340 to 330 K), but it does not suppress the transition completely [8].

Cathodic polarization does not influence this behaviour: the value of $T_t$ and the parameters of the hysteresis loop remain the same as in figure 1(b), and only a high enough cathodic current density and polarization time lead to suppression of the transition. For instance, no MIT is observed in the samples subject to such a treatment at $j = 1$ mA·cm$^{-2}$ for $t = 160$ sec. (figure 1(a), curve 2); on the other hand, if the treatment time is 60 sec. at that same current density, the shift of the $R(\lambda)$ curve (figure 1 (a)) is a little bit smaller than that for the previous case, albeit quite noticeable, and the $R(T)$ dependence is similar to that in figure 1(b).

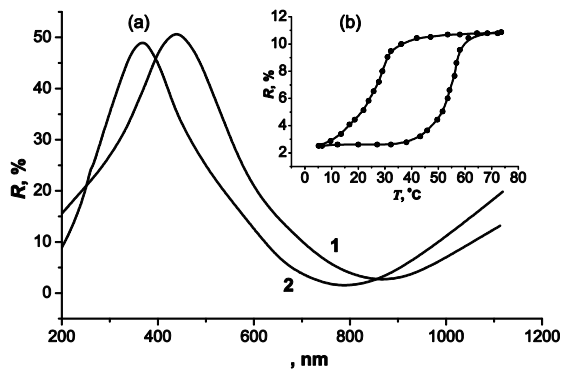

**Figure 1. (a) Reflectance spectrum of the initial (1) and cathodically-treated at $j = 1$ mA·cm$^{-2}$ and $t = 160$ sec. (2) vanadium anodic oxide film, $d = 77$ nm. (b) $R(T)$ variation of $VO_2$ at $\lambda = 900$ nm (both the initial film, and the film cathodically-treated at $j = 1$ mA·cm$^{-2}$ and $t = 60$ sec.).**

The hydrogen concentration (i.e. $x$ in the formula $H_xVO_2$) is straightforward to calculate from the above-presented data on the basis of the Faraday law taking into account the 100% efficiency [3] of the reaction of H$^+$ insertion from the electrolyte into $VO_2$:

$$x = \frac{j \cdot t \cdot \mu}{d \cdot e \cdot \rho \cdot N_A} = 1.57 \quad (1)$$

where $j = 1$ mA·cm$^{-2}$, $t = 60$ sec.; $\mu = 83$ g·mol$^{-1}$ and $\rho = 4.34$ g·cm$^{-3}$ are, respectively, the molecular weight and the density of vanadium dioxide; $d = 77$ nm, $e = 1.6 \cdot 10^{-19}$ C, and $N_A = 6.02 \cdot 10^{23}$ mol$^{-1}$. This value of $x$ is comparable with the data reported in the work [3] for the electrochromic effect in amorphous $V_2O_5$ (where $x = 2$ per one vanadium atom), though it is much higher than that resulting in complete suppression of the MIT in crystalline $VO_2$, which has been estimated to be $x = 0.04$ [6]. As was shown above, for amorphous $VO_2$, such a complete suppression of the MIT took place at $x>>2$ (i.e. after cathodic polarization for $t = 160$ sec. at $j = 1$ mA·cm$^{-2}$). This treatment lead to metallization of the film showing itself in a sharp decrease of two-contact resistance and in a colour change (figure 1); also, the film became insoluble in water, while initial anodic $VO_2$ could be etched out from a substrate in aqueous solutions and even in pure water [8].

It is known that hydrogen introduced into a transition metal oxide acts as a donor impurity [1-4]. In $VO_2$, extra free electrons lead to a lowering of $T_t$ (according to the electronically-driven MIT mechanism) [10] and, finally, to suppression of the transition under cathodic polarization of crystalline vanadium dioxide [7]. On the other hand, in anodic $VO_2$, the appearance of additional delocalized electrons due to doping with donors is impeded because of a high density of localized electronic states playing the role of trap centres in the amorphous material. This accounts for the insensibility of the MIT in anodic vanadium dioxide films to hydrogen insertion up to higher concentrations, as compared to crystalline $VO_2$.

The electrical properties of hydrogen-saturated vanadium anodic oxide films also differ from the behaviour, which is characteristic of the initial (pure) anodic oxide. The latter demonstrates electrical switching with S-shaped $I$-$V$ characteristics due to the MIT in a narrow channel of crystalline $VO_2$ formed inside the film under preliminary electroforming [8]. Another type of current instability and $I$-$V$ curve nonlinearity is exhibited by the films subject to cathodic polarization – see figure 2. In these experiments we have used the anodic oxide films obtained under galvanostatic conditions at anodizing voltage of 45 V ($d = 150$ nm). The phase composition of such samples corresponds to a mixture of vanadium oxides, predominantly to $VO_2$ with a layer close to the $V_2O_5$ stoichiometry at the surface [8,9]. Cathodic polarization at $j = 1$ mA·cm$^{-2}$ for $t = 10$ sec. leads to a partial saturation of the film by hydrogen (equation (1) yields the average value of $x \sim 0.13$ in this case), and the mechanism of nonlinear conductivity can be associated with a redistribution of H$^+$ ions inside the film; a similar process of hydrogen redistribution occurs in $V_2O_5$-gel based hydrated vanadium oxide at the internal electrochromic effect [5].

The order of cycling is shown by numbers in figure 2 (1-2-1-3-4-5-1-6-7-1-8-1), and one can see that there are two steady branches in the $I$-$V$ curve (5-1-2 and 7-1-8) demonstrating reversible switching with the memory effect. It should be noted that the current value is directly proportional to the top Au-electrode surface area.

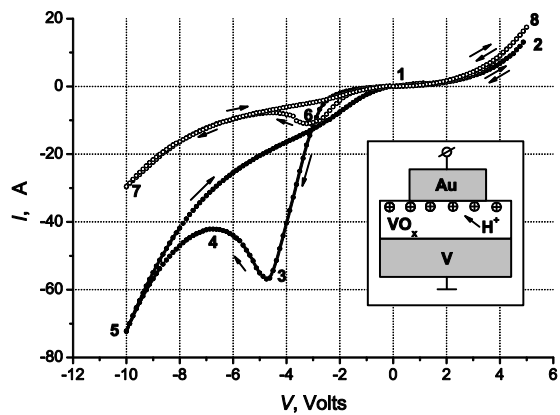

**Figure 2.** Dynamical *I-V* characteristic of the sandwich structure "V/hydrated amorphous vanadium oxide/Au". The structure is schematically depicted in the insert; Au electrode dimensions are 50×50 μm. The oxide film thickness is $d$ = 150 nm, and the cathodic polarization regimes are: $j$ = 1 mA·cm$^{-2}$, $t$ = 10 sec.

We suppose that, originally, the structure is in a high-conductance (albeit non-metallic) state (3) due to the presence of a relatively thick $H_xVO_2$ layer. At a high enough negative bias (minus on the top Au electrode), hydrogen spills over toward the surface $V_2O_5$ layer, the conductivity decreases (note that in $H_xV_2O_5$ a transition to metallic behaviour does not occur even at the highest attainable $x$ values [1]), and the structure passes into a low-conductance state (5-1 and 7-1) via an N-type negative resistance region (4 and 6).

The behaviour described is quite reproducible, the two distinct states (5-1-2 and 7-1-8) are stable with the life-time of more than 30 minutes, and the transition between them occurs via the positive bias.

An analogous process of the hydrogen motion inward/outward across the film has also been discussed in [11] where anodic and cathodic polarizations of vanadium in an electrolyte have been studied. On the other hand, in electrochemical systems, such current peaks (figure 2) are observed in cyclic voltammetry measurements. For example, in the work [12], anodic and cathodic peaks corresponding to the processes of $V^{4+}/V^{5+}$ oxidation/reduction have been observed in the range −1.0 to +1.75 V. In our case (i.e. in a dry system), higher values of the threshold voltage (~ − 4 V – see figure 2) may be ascribed to a voltage drop on series resistances of contacts and transition layers (non-hydrated anodic film).

Certainly, the above speculations are only qualitative, and some additional experimental data are required for the detailed switching mechanism to be fully understood. Particularly, the contributions of ionic currents and H$^+$ concentration gradients should be taken into consideration.

## 4. Conclusion

In conclusion we note that the results presented are of interest from the viewpoint of investigation of low-temperature properties of metallic vanadium dioxide (since the hydrogen insertion stabilizes the metal phase [6]) which would represent a strongly correlated metal prone to MIT. Also, electrical switching in hydrogen-doped vanadium anodic oxide provides the possibility for practical applications of the described structures as electrochemical memory devices [12, 13].

**Acknowledgments**

This work was supported by the Federal Agency for Science and Innovation of the Russian Federation (contract 02.513.11.3351), the Ministry of Education of the Russian Federation and the U.S. Civilian Research and Development Foundation (CRDF Award No. Y5-P-13-01), and the Svenska Institutet (Dnr: 01370/2006).

**References**

[1] Cox P A 1992 *Transition Metal Oxides. An Introduction to their Electronic Structure and Properties* (Oxford: Clarendon)
[2] Aegerter M A, Avellandera C O, Pawlica A, Atic M 1997 *J. Sol-Gel Sci. Technol.* **8** 689
[3] Ord J L, Bishop S D and De Smet D J 1991 *J. Electrochem. Soc.* **138** 208
[4] Gavrilyuk A I 1998 *Ionics* **4** 372
[5] Pergament A L, Kazakova E L, Stefanovich G B 2002 *J. Phys. D: Appl. Phys.* **35** 2187
[6] Andreev V N, Kapralova V M, Klimov V A 2007 *Phys. Solid State* **49** 2318
[7] Chenevas-Paule A 1976 *J. de Phys.* **37** C4-76
[8] Stefanovich G B, Pergament A L, Velichko A A, Stefanovich L A 2004 *J. Phys.: Condens. Matter* **16** 4013
[9] Pergament A L, Stefanovich G B 1998 *Thin Solid Films* **322** 33
[10] Pergament A 2003 *J. Phys.: Condens. Matter* **15** 3217
[11] Clayton G C, De Smet D J 1976 *J. Electrochem. Soc.* **123** 174
[12] Demets G J F, Anaissi F J, Toma H E 2000 *Electrochim. Acta* **46** 547
[13] Liang X F, Chen Y, Shi L et al 2007 *J. Phys. D: Appl. Phys.* **40** 4767.